\documentstyle[aps,prb,twocolumn,floats,graphicx]{revtex}

\begin{document}


\title{Multiple solutions of CCD equations for PPP model of benzene}
\wideabs{
\author{Rafa{\l}  Podeszwa and Leszek Z. Stolarczyk} 
\address{Department of Chemistry, University of Warsaw, Pasteura 1, PL-02-093 Warsaw, Poland}
\author{Karol Jankowski and Krzysztof Rubiniec} 
\address{Institute of Physics, Nicholas Copernicus University, Grudzi\c{a}dzka 5, PL-87-100, Toru\'{n}, Poland}
\date{\today} \maketitle

\begin{abstract}

   To gain some insight into the structure and physical significance of the multiple solutions to the coupled-cluster doubles (CCD) equations corresponding to the Pariser-Parr-Pople (PPP) model of cyclic polyenes, complete solutions to the CCD equations for the ${}A^{-}_{1g}$ states of  benzene are obtained  by means of the homotopy method. By varying the value of the resonance integral 
$\beta$ from $-5.0$~eV to $-0.5$~eV, we cover the so-called  weakly, moderately, and strongly correlated regimes of the model.  For each value of $\beta$ 230 CCD solutions are obtained. It turned out, however, that only for a few solutions a correspondence with some physical states can be established. It has also been demonstrated that, unlike for the standard methods of solving CCD equations, some of the multiple solutions to the CCD equations can be attained by means of the iterative process based on Pulay's direct inversion in the iterative subspace (DIIS) approach.

\end{abstract}

}


\section{Introduction} \label{sec:intro}

The understanding of the electronic correlation effects in cyclic polyenes (or $[M]$annulenes, with the chemical formula ${\rm C}_M {\rm H}_M$,  where $M=4m_0+2$, $m_0 = 1, 2, \ldots$)  has attracted considerable interest for many years. To some extent this interest has been caused by the fact that these molecules provide models for extended one-dimensional metallic-like systems which epitomize the difficulties encountered in the description of extended many-electron systems. A very useful tool in these studies proved to be the coupled cluster (CC) method.~\cite{cizek:66} Even its simplest variant, the CC method with double excitations (CCD),~\cite{cizek:66} is able to reproduce most of the electronic-correlation effects. More sophisticated CCSD~\cite{purvis:82}  and CCSD(T)~\cite{raghavachari:89} methods provide often quite accurate approximations to the full configuration-interaction (FCI) results (which define the limit for a given orbital basis set). The CC method furnishes a set of coupled nonlinear inhomogeneous  equations, of unknowns representing the amplitudes of the CC operator $\hat{T}$ (the $t$ amplitudes), in which the number of equations is equal to the number of unknowns. However, due to their nonlinearity, the CC equations have multiple solutions; this problem was first analyzed by \v{Z}ivkovi\'{c} and Monkhorst.~\cite{zivkovic:77,zivkovic:78} It is still little known about the mathematical properties of these solutions. 

Recently,~\cite{kowalski:98a,jankowski:99a,jankowski:99d,jankowski:99e} for the first time, 
some insight into the structure of the complete set of solutions to the CC equations has been gained by applying 
the powerful homotopy (continuation) method~\cite{drexler:78,morgan:87} to the CCD and the CCSD equations corresponding 
to some four-electron four-orbital systems, known as the H4 and P4 models.~\cite{jankowski:80} 
These studies included the correspondence between CC and configuration interaction (CI) methods,\cite{jankowski:99a}
and the influence of the approximate form of the cluster operator on the structure of solutions of the associated equations.\cite{jankowski:99d} 
Within the symmetry and spin-adapted CCD formalism, the CCD equations for the H4 model comprise a system of six coupled quadratic inhomogeneous equations for six unknowns (the $t_2$ amplitudes of the double-excitation operator $\hat{T}_2$). According to the B\'{e}zout theorem, the maximum number of solutions amounts in this case to $2^6 = 64$. Yet Kowalski and Jankowski~\cite{kowalski:98a} showed that  the CCD equations for the H4 model case have only 12 solutions (6 real, and 3 pairs of complex solutions). Some of these solutions corresponded to the variational results of the configuration-interaction with doubles (CID) method. 

  In the present paper we study, by means of the homotopy method, multiple solutions of the CCD equations for the six-electron six-orbital system corresponding to the Pariser-Parr-Pople (PPP)~\cite{pariser:53a,pariser:53b,pople:53} model of benzene. This system is of special interest for theoretical chemists: the sextet of $\pi$-electrons is responsible for the peculiar chemical properties of benzene which is the archetypal aromatic molecule. The PPP model is known to provide a simplified yet reliable picture of many-electron systems.~\cite{DelRe:90}  For benzene, one gets an almost perfect description of the $\pi$-electron part of the excitation spectrum by optimizing the empirical parameters of the PPP model at the FCI level.~\cite{karwowski:73a,karwowski:73b} Moreover, in the PPP model the strength of the electronic-correlation effects can be controlled without changing the molecular geometry, by simply adjusting the value of the so-called resonance integral $\beta$ $(< 0)$. Very high symmetry of the PPP benzene model ensures that the number of symmetry-independent parameters entering the exact (FCI) or an approximate (e.g., CCD) wave function is small. All that makes the PPP benzene model particularly suitable for testing quantum-chemical methods dealing with the electronic correlations.

  Benzene is the first member of the $[M]$annulene family for which, as mentioned above, the understanding of the electronic structure is of rather general significance. These systems, especially when described within the PPP model, seem deceptively simple: their one-electron states are fully determined by the spatial symmetry, and so is their Hartree-Fock (HF) determinantal wave function. Yet  $[M]$annulenes turn out to be a challenge to the existing computational methods of treating the electronic correlations: it has been found that for larger $[M]$annulenes (corresponding to $m_0 > 2$), in the so-called strong correlation regime of $\beta$, the CCD method breaks down completely, as no real solution of the CCD equations exists for $\beta$ greater than some critical value.~\cite{paldus:84a,paldus:84b} A recent CC study of $[M]$annulenes, taking into account the double (D), triple (T), and quadruple (Q) excitations in the CC operator, showed that even the CCDTQ method breaks down for these systems when the correlation effects become sufficiently strong.~\cite{podeszwa:02a}  For the PPP  benzene model, a real CCD solution representing the ground state can be found for any $\beta \leq 0 $. However, as $\beta$ approaches $0$, the CCD $t_2$ amplitudes   deviate more and more from the corresponding FCI values, thus suggesting that some correlation effects peculiar to the higher $[M]$annulenes may surface already in the strongly correlated regime of the PPP benzene model. This provides an additional motivation for the present study.

\section{PPP model of benzene}\label{sec:PPP}

 A detailed description of the PPP model of annulenes may be found in Ref.~\onlinecite{podeszwa:02a}.
 In benzene the C atoms form a  regular hexagon, and the C--C bonds are assumed to be of the length $R^0 = 1.4$~\AA. The PPP model invokes the $\pi$-electron approximation and  describes the six $\pi$ electrons of benzene by using a minimal basis set of $2p_z$ atomic orbitals associated with the six carbon atoms. The set of these $\pi$ atomic orbitals ($\pi$AOs) is then subject to the symmetrical orthonormalization procedure of L\"{o}wdin,~\cite{lowdin:50} yielding the set of six orthonormalized $\pi$ atomic orbitals ($\pi$OAOs), denoted by $\chi_m \,$, $m = 0, \pm 1, \pm 2, 3$ (we use here the numbering system employed in Ref.~\onlinecite{podeszwa:02a}). The Fock-space Hamiltonian $\hat{H}$ for benzene, built according to the prescriptions of the PPP model, is given in Eq.~(2) of Ref.~\onlinecite{podeszwa:02a}. The following semiempirical parameters are used in the PPP model: \\
(i) $\alpha$, the so-called Coulomb integral for the carbon atom,
representing the binding energy of electron described by the $\pi$OAO $\chi_m \,$ (for simplicity we put $\alpha = 0$~eV), \\
(ii) $\beta (< 0)$, the resonance integral, corresponding to the electron transfer between the neighboring $\pi$OAOs, $\chi_m$ 
and $\chi_{m+1} \,$ (our results are for $\beta = -5.0$, $-4.0$, $-3.0$, $-2.5$, $-2.0$, $-1.5$, $-1.0$, and $-0.5$~eV), \\
(iii) $\gamma_{mn}$, representing the two-center two-electron integrals 
$\langle \chi_m \chi_n |\chi_m \chi_n \rangle \,$; usually one calculates  $\gamma_{mn} = \gamma( R_{mn})$, where $ R_{mn}$ is the distance between the centers of orbitals $\chi_m$ and $\chi_n \,$  and function $\gamma(R)$ is given by some simple analytical formula. We use the Mataga-Nishimoto formula~\cite{mataga:57}, $\gamma(R) = e^2 [R + e^2 (\gamma^{\rm o})^{-1}]^{-1} \,$, where $e$ is the electron charge and $\gamma^0 = \gamma(0) = 10.84$~eV. 

  The point-symmetry group of benzene is ${\rm D}_{6{\rm h}}$, but its subgroup ${\rm C}_6$ is sufficient for the symmetry considerations in the $\pi$-electron approximation. The molecular orbitals of the $\pi$ symmetry ($\pi$MOs),  expressed as linear combinations of $\pi$OAOs,  are completely determined by the projections onto the irreducible representations of the ${\rm C}_6$ group and read as
\begin{eqnarray}\label{eq:MOs}
\psi_k &=&  6^{-1/2} \big[ \chi_0 + e^{ k \pi i /3} \chi_1 + e^{- k \pi i /3} \chi_{-1} \nonumber \\
&&{}+ e^{2k \pi i /3} \chi_2 + e^{-2k \pi i /3} \chi_{-2} + e^{ k \pi i } \chi_3 \big] \,,
\end{eqnarray}
where $k = 0, \pm 1, \pm 2, 3$ stand for the symmetry labels. For $k \neq 0,3$ the $\pi$MOs are complex, $\psi_k^* = \psi_{-k}$, and, due to the time-reversal symmetry, correspond to degenerate orbital energies, $\epsilon(k) = \epsilon(-k)$. In the restricted Hartree-Fock (RHF) description of the ground state of benzene the occupied $\pi$MOs correspond to $k = 0, \pm 1$ and the unoccupied $\pi$MOs to $k = \pm 2, 3$. The formulas for the orbital energies and the HF $\pi$-electron energy read as~\cite{podeszwa:02a}
\begin{equation}\label{eq:e(k)}
\epsilon(k) = 2 \beta \cos (k \pi /3) + \gamma^0 - [v(k ) + v(k+1) + v(k-1)] \,,
\end{equation}
\begin{equation}\label{eq:EHF}
E_{\rm HF} = 8 \beta  +  3 \gamma^0 - [3 v(0) +4 v(1) + 2 v(2)] \,.
\end{equation}
These formulas depend on parameters $\beta$, $\gamma^0$, and the two-electron integrals calculated in the $\pi$MO basis:
\begin{eqnarray}\label{eq:v(q)}
v(q)  &=& v(-q) = \langle \psi_{k_1 + q} \psi_{k_2 - q} | \psi_{k_1}
\psi_{k_2} \rangle  \nonumber \\
&=& 6^{-1}   \big[ \gamma^0 + 2 \cos (q \pi /3) \gamma (R^0) \nonumber \\ 
& &{}+ 2 \cos (2 q \pi /3) \gamma (\sqrt{3}\, R^0) + \cos (q \pi) \gamma (2R^0) \big] \,, 
\end{eqnarray}
where $q = 0, 1, 2, 3$, and the modulo-6 rule of addition is assumed for the symmetry labels.

\section{CCD method for the PPP model of benzene}\label{sec:CC}

In the standard single-reference CC theory, the ground-state FCI wave function $\Psi$ for a six-electron system is represented as
\begin{equation}\label{eq:expT}
\Psi = \exp(\hat{T}) \Phi \,,
\end{equation}
where $\hat{T} = \hat{T}_1+\hat{T}_2+\cdots+\hat{T}_6$ is the CC operator and $\Phi$ is the RHF determinantal wave function, playing the role of the reference configuration. The $\hat{T}_n$ components of the CC operator correspond to the connected $n$-tuple excitations from occupied to unoccupied spin-orbitals. Each $\hat{T}_n$ operator depends on some linear parameters, hereafter referred to as the $t_n$ amplitudes; the ordered set of all $t_n$ amplitudes form a vector denoted by ${\bf t}_n$. In the CC theory one introduces a similarity-transformed Hamiltonian, $\hat{\overline{H}} = \exp(-\hat{T}) \hat{H} \exp(\hat{T})$, whose amplitudes $\bar{h} ^{pq\ldots}_{rs\ldots}$
are certain connected functions of the amplitudes of the Fock-space  Hamiltonian, and the $t_n$ amplitudes. The electronic correlation energy for the ground state $\Psi$ may be calculated in the CC theory as
\begin{equation}\label{eq:Ecorr}
E_{\rm corr} = \langle\Phi|\hat{\overline{H}}|\Phi\rangle - E_{\rm HF} =
\bar{h}({\bf t}_1, {\bf t}_2) - E_{\rm HF} \,,
\end{equation}
i.e., it is a function of only $t_1$ and $t_2$ amplitudes. The $t_n$ amplitudes ($n =1,2,\ldots,6$) may be calculated by solving the set of CC equations:
\begin{mathletters}
\label{eq:CC}
\begin{eqnarray}\
\langle\Phi^a_i|\hat{\overline{H}}|\Phi\rangle = & \bar{h}^a_i({\bf t}_1,
{\bf t}_2, {\bf t}_3) & = 0 \,,  \label{eq:CCS} \\
\langle\Phi^{ab}_{ij}|\hat{\overline{H}}|\Phi\rangle = & \bar{h}^{ab}_{ij}
({\bf t}_1, {\bf t}_2, {\bf t}_3, {\bf t}_4) & = 0 \,, \label{eq:CCD} \\
& \vdots & \nonumber \\
\langle\Phi^{abcdef}_{ijklmn}|\hat{\overline{H}}|\Phi\rangle = &
\bar{h}^{abcdef}_{ijklmn}
({\bf t}_1, {\bf t}_2, {\bf t}_3, {\bf t}_4, {\bf t}_5, {\bf t}_6 ) \mbox{ } & = 0 \label{eq:CCH}
\,,
\end{eqnarray}
\end{mathletters}
where $\Phi^{ab\ldots}_{ij\ldots}$ is an $n$-tuply excited configuration. Written in an explicit form, the CC equations (\ref{eq:CC}) form a set of coupled inhomogeneous nonlinear equations, with the number of unknowns (the $t_n$ amplitudes) equal to the number of equations.  On a basis of the FCI method, it can be shown that  the {\em exact\/}  $t_n$ amplitudes are real. In the simplest approximate variant of the CC method, the CCD one, one puts $\hat{T} = \hat{T}_2$ and neglects $t_1$, $t_3$, and $t_4$ amplitudes in Eqs.~(\ref{eq:Ecorr}) and (\ref{eq:CCD}) (as well as the remaining CC equations). The CCD equations (\ref{eq:CCD}) then become a set of coupled inhomogeneous quadratic equations for the unknown $t_2$ amplitudes, and from Eq.~(\ref{eq:Ecorr}) an approximate correlation energy, $E_{\rm corr} ^{\rm CCD}$, is calculated.

In the PPP model of benzene the occupied and unoccupied orbitals belong to different representations of the ${\rm C}_6$ group, which causes $\hat{T}_1$ and $\hat{T}_5$ to vanish by symmetry. Because ${\bf t}_1 = {\bf 0}$, the RHF function $\Phi$ becomes equal to the Brueckner determinantal function.~\cite{paldus:73} In this case Eqs.~(\ref{eq:CCS}) are automatically satisfied, and the CCD method becomes equivalent to the CCSD one. When the nonorthogonal spin adaptation of the CCD equations is performed (see, e.g., Ref.~\onlinecite{stolarczyk:84}), the spin-adapted $t_2$ amplitudes for benzene may be written as $t(k_1,k_2,q)$,  where  $k_1$ and $k_2$ are occupied-$\pi$MOs labels, and $q$ $(\geq 0)$ is chosen such that $k_1 + q$ and $k_2 + q$  are unoccupied-$\pi$MOs labels (see Ref.~\onlinecite{podeszwa:02a}). It can be shown that there are 11 different sets of $k_1,k_2$, and $q$, thus defining 11 $t_2$ amplitudes for benzene. By assuming that these amplitudes are real and employing the time-reversal symmetry, one finds a symmetry constraint~\cite{podeszwa:02a}
\begin{equation}\label{eq:t-rev}
t(k_1, k_2, q) = t(-k_2, -k_1,q) \,,
\end{equation}
which reduces the number of the symmetry-independent $t_2$ amplitudes for benzene to 8.

  In Ref.~\onlinecite{podeszwa:02a} we employed the general non-orthogonally spin-adapted CCD equations for the PPP model of the $[M]$annulenes, with the $t_2$ amplitudes $t(k_1, k_2, q)$ subject to the symmetry constraint (\ref{eq:t-rev}). For benzene, the set of these equations may be written as
\begin{equation}\label{eq:CCD2}
a_i + \sum_{j=1}^{8} b_{ij} x_j + \sum_{j=1}^{8} \sum_{k=1}^{8}c_{ijk} x_j x_k = 0 \,, 
\end{equation}
where $i = 1,2, \ldots, 8$, and the unknowns $x_j$, $j = 1,2, \ldots, 8$ stand for the symmetry-independent $t_2$ amplitudes, see Table~\ref{tab:indices}. The formula for the electronic-correlation energy now reads as
\begin{equation}\label{eq:Ecorr2}
E_{\rm corr}^{\rm CCD} = \sum_{j=1}^{8} d_j x_j \,. 
\end{equation}
The linear coefficients in Eqs.~(\ref{eq:CCD2}) and~(\ref{eq:Ecorr2}): $a_i$, $b_{ij}$ (for $i \neq j$), $c_{ijk} = c_{ikj}$, and $d_j$ can be expressed as some linear combinations of the two-electron integrals defined in Eq.~(\ref{eq:v(q)}). Only the diagonal elements $b_{ii}$ depend on parameter $\beta$:
\begin{equation}\label{eq:bii}
b_{ii} = \Delta e_i - 2 v(0) \,,
\end{equation}
where
\begin{equation}\label{eq:Deltaei}
\Delta e_i = \Delta e(k_1,k_2,q) = \epsilon(k_1 + q) +  \epsilon(k_2 - q)
- \epsilon(k_1) - \epsilon(k_2) \,,
\end{equation}
through the dependence on $\beta$ of the orbital energies (\ref{eq:e(k)}). There is still some symmetry hidden in the set of CCD equations (\ref{eq:CCD2}): due to the so-called alternancy symmetry of the PPP Hamiltonian $\hat{H}$ (see Ref.~\onlinecite{koutecky:85}, and references therein) one finds that
\begin{equation}\label{eq:altsym}
x_2 = t(0,0,2) = t(1,-1,2) = x_3 \,,
\end{equation}
(for the general formula for [M]annulenes, see Ref.~\onlinecite{podeszwa:02a}). This property propagates into Eq.~(\ref{eq:CCD2}) making them invariant with respect to the interchange of indices $2$ and $3$. However, our set of CCD equations is not {\em explicitly\/} adapted to the alternancy symmetry, and solutions breaking this symmetry are, in principle, possible.

\begin{table}[tbp]
\caption{Correspondence between indices of Eq.~(\ref{eq:CCD2}) and quasimomentum indices $k_1$, $k_2$, and $q$.}
\label{tab:indices}
\begin{tabular}{lrrr}
$i $ & $k_1$ & $k_2$ & $q$ \\
\hline
1   & 1 & $-1$ & 1 \\
2   & 0 &  0 & 2 \\
3   & 1 & $-1$ & 2 \\
4   & 1 &  0 & 2 \\
5   & $-1$&  1 & 3 \\
6   & 0 &  0 & 3 \\
7   & 0 &  1 & 3 \\
8   & 1 &  1 & 3 \\
\end{tabular}
\end{table}

The usual method of solving the CCD equations is based on an iterative procedure, which in the case of Eq.~(\ref{eq:CCD2}) may be written as follows:
\begin{eqnarray}\label{eq:CCD3}
x _i^{(n+1)} &=& - (\Delta e_i)^{-1} \big[ a_i + \sum_{j=1}^{8} (b_{ij} - \Delta e_i \delta_{ij}) x_j^{(n)} \nonumber \\
& & {}+ \sum_{j=1}^{8} \sum_{k=1}^{8}c_{ijk} x_j^{(n)} x_k^{(n)} \big] \,, 
\end{eqnarray}
with $x _i^{(0)} = 0$, $i=1,2, \ldots,8$. The first iteration furnishes $t_2$ amplitudes that substituted into Eq.~(\ref{eq:Ecorr2}) give the second-order M{\o}ller-Plesset (MP2) correlation energy. When convergent, this simple iterative procedure provides a {\it single} solution to the CCD equations (\ref{eq:CCD2}). Such a solution is bound to be real and to preserve the alternancy symmetry, which corresponds to the fulfillment of Eq.~(\ref{eq:altsym}). In our calculations for the PPP model of benzene with $\beta \in [-5.0$~eV,$0$~eV$]$, we found that the above described simple iterative procedure is indeed convergent, and the correlation energy calculated by substituting the convergent $t_2$ amplitudes into Eq.~(\ref{eq:Ecorr2})  approximates the FCI value for the ground state. The agreement between the CCD and FCI results ($t_2$ amplitudes and $E_{\rm corr}$) is very good in the weakly and moderately correlated regimes ($\beta$ in the vicinity of $-5.0$~eV and $-2.5$~eV, respectively), but becomes rather poor in the strongly correlated regime ($\beta > -0.5$~eV). The CCD and FCI results for $\beta = -2.5$~eV and $\beta = -0.5$~eV may be found in Ref.~\onlinecite{podeszwa:02a}.

\section{Multiple solutions of CCD equations for benzene}\label{sec:sol}

\begin{table*}[tb!]
\caption{Numbers of states of different symmetry for various $\beta$ given in eV.}
\label{tab:number_of_states}
\begin{tabular}{lrrrrrrrr}
			 & $-5.0$&$-4.0$&$-3.0$&$-2.5$&$-2.0$&$-1.5$&$-1.0$&$-0.5$ \\
\hline
real, symmetric		 &    8  &    8 &    6 &    6 &    6 &    8 &    4 &	6 \\
real, broken symmetry	 &   24  &   20 &   20 &   24 &   24 &   22 &   20 &   16 \\
complex, symmetric	 &  116  &  116 &  118 &  118 &  118 &  116 &  120 &  118 \\
general complex, broken symmetry &   80  &   84 &   84 &   80 &   80 &   84 &   84 &   88 \\

special complex, broken symmetry &    2  &    2 &    2 &    2 &    2 &    0 &    2  &    2 \\
total			 &  230  &  230 &  230 &  230 &  230 &  230 &  230 &  230 \\
\end{tabular}
\end{table*}

The CCD equations (\ref{eq:CCD2}) comprise a set of 8 coupled quadratic 
inhomogeneous  equations (with real coefficients)  for 8 unknowns. 
According to the B\'{e}zout theorem, such equations may have up to
$2^8 = 256$ solutions, complex in general. In principle, a {\em complete\/} set of solutions can be found by means of 
the homotopy (continuation) method.~\cite{drexler:78,morgan:87} 
Below we present the results obtained by applying the homotopy method to the CCD equations (\ref{eq:CCD2}) corresponding to various values 
of the resonance integral $\beta$: from $-5.0$~eV (representing the weakly correlated regime) to $-0.5$~eV (representing the strongly correlated regime). The FCI results used for comparison were calculated with {\sc gamess}.\cite{schmidt:93}

  Equations (\ref{eq:CCD2}) have been derived by taking into account the 
spin and the time reversal symmetries, as well as the reality of the 
$t_2$ amplitudes, see Eq.~(\ref{eq:t-rev}). However, these equations 
may have also {\em complex\/} solutions. Such solutions have to appear 
in pairs: if ${\bf x} = (x_1, x_2, \ldots, x_8)$ is a complex solution,  
then its complex-conjugate ${\bf x}^*$ is also a solution. For a complex solution, the correlation energy calculated from Eq.~(\ref{eq:Ecorr2}) assumes (in general) a complex value, and the complex-conjugate solutions correspond to the complex-conjugate values of the correlation energy. Pairs of complex solutions, ${\bf x}$ and ${\bf x}^*$, will be called degenerate, since they correspond to the same 
real part of the complex correlation energy calculated 
from Eq.~(\ref{eq:Ecorr2}). Some solutions may also violate the equality (\ref{eq:altsym}) derived from the alternancy-symmetry. Such symmetry-broken solutions, 
for which $x_2 \neq x_3$, also have to appear in pairs: if ${\bf x}$ 
is such a solution, then ${\bf x}'$, in which the values  $x_2$ and $x_3$ are interchanged, has to be a solution as well.  Pairs of the real symmetry-broken solutions, ${\bf x}$ and ${\bf x}'$, are also degenerate. In the case of the general complex symmetry-broken solutions, ${\bf x}$ and ${\bf x}^*$, ${\bf x}'$ and ${\bf x}'^*$ form 
a degenerate quadruplet. We have found, however, a special class of complex symmetry-broken solutions in which the only complex values correspond to $x_2 = x_3^*$. In such a case one has ${\bf x}^* = {\bf x}'$, and the pair of solutions ${\bf x}$ and ${\bf x}'$ corresponds to the same {\em real\/} value of the correlation energy calculated from Eq.~(\ref{eq:Ecorr2}).  Thus, the solutions of Eqs. (\ref{eq:CCD2}) may be classified into five distinct categories: real symmetric (which are non-degenerate), real symmetry-broken, complex symmetric, general complex symmetry-broken, and a special class of complex symmetry-broken with the real energy.

In Table~\ref{tab:number_of_states} the number of  CCD solutions belonging to different categories is presented for several  values of the resonance integral $\beta$. The total number of solutions (230) is surprisingly large: it is only slightly smaller than the B\'{e}zout upper bound (256), and much larger than the number of solutions for the H4 model (6 real and 6 complex, compared to the upper limit 64 allowed by the B\'{e}zout theorem\cite{kowalski:98a}). 
The number of CCD solutions is also much larger than the number of the FCI solutions for the PPP model of benzene, having the same symmetry as the RHF wavefunction $\Phi$, see further discussion. This implies that most of the CCD solutions have no physical meaning. Since the FCI method is equivalent to the full coupled-cluster (FCC) method, it is interesting that the CCD equations, which have smaller order and a smaller number of the unknowns, have more solutions than the more sophisticated FCC equations. This suggests that the truncated CC equations are unable to utilize all the symmetries that are present in the many-electron Hamiltonian. For the H4 model it was observed~\cite{jankowski:99d} that the number of solutions increased from CCD ($12$ solutions) to CCSD ($60$ solutions), and then decreased to $7$ for FCC.

While for the H4 model the number of pairs of complex solutions was equal to the number of real solutions, for the PPP model of benzene there are much more complex than real solutions. 
Since all the FCI results are real, the complexity of the solutions must be caused by the truncation
of the CC operator. Since all the linear coefficients in the CCD equations (\ref{eq:CCD2}) 
are real, each complex solution must have its complex conjugate counterpart. 
However, while solving these equations by using the homotopy method, 
we have found in several instances that some complex-conjugate solutions were missing 
(duplicated solutions were also encountered). Since the problem has not been previously reported,
we think that it may have arisen due to the numerical complexity of the problem 
(a lot of roots to be traced in the homotopy algorithm). 
It brings in the question whether the CCD solutions presented here are {\em complete\/}. 
After removing duplicates and adding solutions that must be present due to symmetry, 
we have found that the total number of states is constant (equal to $230$) for each considered value of $\beta$. 
The number of solutions preserving the alternancy symmetry has also appeared to be constant (equal to $124$). 
It seems therefore unlikely that certain states has been overlooked.

Due to the alternancy symmetry of the PPP model, the six-electron symmetry states of benzene split into two categories, 
denoted by ``minus'' and ``plus.''~\cite{pariser:56} Including the spin symmetry and the spatial symmetry 
of the ${\rm D}_{6{\rm h}}$ group, the ground-state RHF wave function  $\Phi$ corresponds to the symmetry 
label ${}A^{-}_{1g}$. Among $400$ FCI states of benzene generated by {\sc gamess}\cite{schmidt:93} 
there are only 18 states corresponding to the ${}A^{-}_{1g}$ symmetry. 
These  ${}mA^{-}_{1g}$-states ($m =1,2,\ldots,18$) are, in general, 
non-orthogonal to $\Phi$, and thus may be expressed in the form of the CC expansion (\ref{eq:expT}). 
For each case, the corresponding $t_n$ amplitudes may be extracted from the FCI linear coefficients:
there are $7$ $t_2$ amplitudes, $2$ $t_3$ amplitudes, $7$ $t_4$ amplitudes, 
and only a single $t_6$ amplitude (only non-redundant parameters are counted).

We are interested in identifying solutions to the CCD equations which have physical significance, i.e., which correspond to some states of the ${}^1A^{-}_{1g}$ characteristic represented in the model by relevant solutions to the FCI equations. In order to gauge a similarity between the $t_2^{\rm CCD}$ amplitudes corresponding to a given CCD solution and the $t_2^{\rm FCI}$ amplitudes corresponding to some FCI solution of the ${}A^{-}_{1g}$ symmetry, we use  parameters $\theta$ and $\eta$ defined below:
\begin{equation}\label{eq:theta,eta}
\theta = \arccos \left( \frac {  {\bf t}_2^{\rm CCD} {\bf t}_2^{\rm FCI}}
{| {\bf t}_2^{\rm CCD}| |{\bf t}_2^{\rm FCI}|} \right), \qquad
\eta = \frac{|{\bf t}_2^{\rm CCD}|}{|{\bf t}_2^{\rm FCI}|} \, ,
\end{equation}
where we use the real part of the amplitudes ${\bf t}_2^{\rm CCD}$. Here $\theta$ measures the angle between the vectors, and $\eta$---the ratio of the vector lengths; in the above analysis we use the vectors corresponding to the full set of $t_2$  
amplitudes (of the dimension 11), i.e., containing the symmetry-redundant amplitudes fulfilling Eq.~(\ref{eq:t-rev}). In principle, 
a complex solution may be considered an approximation to some real solution of the FCI equations as long as the imaginary 
parts of the $t_2$ amplitudes are small in comparison to the real parts.

Some of the CCD solutions, obtained for several values of $\beta$, are characterized in Table~\ref{tab:theta,eta}. Solution numbers $n_{\rm CCD}$ are assigned in accordance with the increase (of the real part) of the corresponding correlation energy value calculated from Eq.~(\ref{eq:Ecorr2}). For instance, $n_{\rm CCD}=005$ denotes the 5th solution. Due to the large number of solutions, we consider only those that are the closest to some FCI ones,  i.e. those corresponding  to $\theta$ and $\eta$  closest to $0$ and $1$, respectively. (The complete set of solutions may be obtained from the authors).  

\begin{table*}[t]
\caption{Correspondence between some states of the PPP model of 
benzene and solutions to the CCD equations for various $\beta$ values.
FCI and CCD correlation energies are in eV, $n_{\rm CCD}$ is
CCD solution number, $\theta$ (in rad) and $\eta$ parameters are defined in Eq.~(\ref{eq:theta,eta}).} 
\label{tab:theta,eta}
\begin{tabular}{rrrr@{}lrr}
\multicolumn{1}{c}{Solution}  & \multicolumn{1}{c}{$E_{\rm FCI}$} & \multicolumn{1}{c}{$n_{\rm CCD}$} &  
   \multicolumn{2}{c}{$E_{\rm CCD}$} & \multicolumn{1}{c}{$\theta$}  &   \multicolumn{1}{c}{$\eta$}     \\ 
\multicolumn{1}{c}{characteristic} & \multicolumn{6}{c}{}\\													    
\hline
\multicolumn{7}{c}{$\beta=-0.5\mbox{ eV}$}\\
$1\, {}^1\!A_{1g}^{-} $  & $-5.389\mbox{ }786 $  & $ 005$       &  $-8.$&$290\mbox{ }596$   &$ 0.2430 $  & $1.6557$ \\ 
$10\, {}^1\!A_{1g}^{-} $  & $ 6.914\mbox{ }750 $  & $ 138,139 $  &  $7.$ &$154\mbox{ }622 \pm i 0.234\mbox{ }294$   &$ 0.3395 $  & $ 0.9860$   \\ 
$14\, {}^1\!A_{1g}^{-} $  & $ 9.825\mbox{ }925 $  & $ 193     $  &  $11.$& $390\mbox{ }287 $                 &$ 0.5647 $  & $ 1.4070 $  \\  
$18\, {}^1\!A_{1g}^{-} $  & $ 20.406\mbox{ }881$  & $ 200,201 $  &  $ 12.$&$648\mbox{ }277 \pm i 1.072\mbox{ }852 $  &$ 0.4441 $  & $ 1.1293 $  \\
\multicolumn{7}{c}{$\beta=-1.0\mbox{ eV}$}\\
$1\, {}^1\!A_{1g}^{-} $  & $ -3.421\mbox{ }010 $  & $ 005 $     &  $-3.$&$939\mbox{ }967 $   &$ 0.1589 $  & $  1.2101$ \\ 
$10\, {}^1\!A_{1g}^{-} $  & $ 10.422\mbox{ }514 $  & $ 160,161 $  &  $10.$&$802\mbox{ }034 \pm i 1.050\mbox{ }428 $  &$ 0.2776 $  & $ 1.1035 $  \\  
$18\, {}^1\!A_{1g}^{-} $  & $ 24.749\mbox{ }882 $  & $ 206,207 $  &  $20.$&$032\mbox{ }365 \pm i 1.114\mbox{ }519 $  &$ 0.4639 $  & $ 1.0800 $ \\
\multicolumn{7}{c}{$\beta=-1.5\mbox{ eV}$}\\
$1\, {}^1\!A_{1g}^{-} $  & $ -2.330\mbox{ }250 $  & $ 005     $  & $ -2.$&$397\mbox{ }334                $  &$ 0.0851 $  & $  1.0505 $ \\
$11\, {}^1\!A_{1g}^{-} $  & $ 15.466\mbox{ }311 $  & $ 193     $  & $ 17.$&$676\mbox{ }035                $  &$ 0.5495 $  & $  0.6366 $ \\
$14\, {}^1\!A_{1g}^{-} $  & $ 19.687\mbox{ }009 $  & $ 206,207 $  & $ 23.$&$335\mbox{ }177 \pm i 0.877\mbox{ }270 $  &$ 0.4564 $  & $  0.9985 $ \\
$18\, {}^1\!A_{1g}^{-} $  & $ 29.397\mbox{ }385 $  & $ 204,205 $  & $ 21.$&$056\mbox{ }124 \pm i 3.647\mbox{ }862 $  &$ 0.3480 $  & $  1.7948 $ \\ 
\multicolumn{7}{c}{$\beta=-2.0\mbox{ eV}$}\\
$1\, {}^1\!A_{1g}^{-} $  & $ -1.726\mbox{ }025 $  & $ 005     $  & $ -1.$&$729\mbox{ }921  $             &$ 0.0495 $  & $  1.0119 $ \\
$11\, {}^1\!A_{1g}^{-} $  & $ 20.093\mbox{ }376 $  & $ 194     $  & $ 21.$&$991\mbox{ }453    $              &$ 0.5115 $  & $  0.5779 $ \\
$12\, {}^1\!A_{1g}^{-} $  & $ 22.664\mbox{ }453 $  & $ 204,205 $  & $ 24.$&$340\mbox{ }342 \pm i 4.664\mbox{ }894  $  &$ 0.1826 $  & $  2.3779 $ \\
$18\, {}^1\!A_{1g}^{-} $  & $ 34.788\mbox{ }885 $  & $ 210,211 $  & $ 26.$&$939\mbox{ }140 \pm i 1.936\mbox{ }787   $  &$ 0.4914 $  & $  1.0048 $ \\
\multicolumn{7}{c}{$\beta=-2.5\mbox{ eV}$}\\
$1\, {}^1\!A_{1g}^{-} $  & $ -1.363\mbox{ }707 $  & $ 005     $  & $-1.$&$358\mbox{ }839 $                &$ 0.0319 $  & $  1.0017 $ \\
$12\, {}^1\!A_{1g}^{-} $  & $ 27.206\mbox{ }496 $  & $ 201,202 $  & $ 27.$&$653\mbox{ }154 \pm i 5.484\mbox{ }227  $  &$ 0.1826 $  & $  2.7398 $ \\
$18\, {}^1\!A_{1g}^{-} $  & $ 41.714\mbox{ }245 $  & $ 226     $  & $ 70.$&$723\mbox{ }477 $                 &$ 0.3813 $  & $  0.9327 $ \\
\multicolumn{7}{c}{$\beta=-3.0\mbox{ eV}$}\\
$1\, {}^1\!A_{1g}^{-} $  & $ -1.126\mbox{ }551 $  & $ 005     $  & $-1.$&$121\mbox{ }438                $  &$ 0.0222 $  & $  0.9988 $ \\
$12\, {}^1\!A_{1g}^{-} $  & $ 31.696\mbox{ }270 $  & $ 196,197 $  & $30.$&$987\mbox{ }726 \pm i 6.215\mbox{ }649   $  &$ 0.1914 $  & $  2.8482 $ \\
$18\, {}^1\!A_{1g}^{-} $  & $ 49.302\mbox{ }346 $  & $ 226     $  & $78.$&$427\mbox{ }410 $                &$ 0.2985 $  & $  1.0682 $ \\
\multicolumn{7}{c}{$\beta=-4.0\mbox{ eV}$}\\
$1\, {}^1\!A_{1g}^{-} $  & $ -0.836\mbox{ }853 $  & $ 007     $  & $ -0.$&$833\mbox{ }688                $  &$ 0.0125 $  & $  0.9979 $ \\
$18\, {}^1\!A_{1g}^{-} $  & $ 64.911\mbox{ }337 $  & $ 226     $  & $ 93.$&$844\mbox{ }690                $  &$ 0.2165 $  & $  1.1505 $ \\
\multicolumn{7}{c}{$\beta=-5.0\mbox{ eV}$}\\
$1\, {}^1\!A_{1g}^{-} $  & $ -0.666\mbox{ }635 $  & $ 005     $  & $  -0.$&$664\mbox{ }763$                 &$ 0.0080 $  & $  0.9982 $ \\
$18\, {}^1\!A_{1g}^{-} $  & $ 80.708\mbox{ }757 $  & $ 226     $  & $109.$&$317\mbox{ }542 $                &$ 0.1762 $  & $  1.1598 $ \\
\end{tabular}
\end{table*}

Solution $005$ for each of the $\beta$ values (except for $\beta=-4.0\mbox{ eV}$, where it is solution $007$) 
is considered to be the ground state,~\cite{paldus:84a,paldus:84b}
and can be obtained by applying the standard iterative process of Eq.~(\ref{eq:CCD3}). Indeed,
this solution is the most similar to the FCI ground state ${}A^{-}_{1g}$ for all the $\beta$ values studied
(in the strongly correlated region the similarity is, however, rather poor~\cite{podeszwa:02a}).
Except for that state, there is little similarity between the CCD  solutions considered and the FCI states, 
both in energy and amplitudes.

 For all values of $\beta$ the state corresponding to  the ground state
has, in fact, the energy {\em larger\/} than some of the other CCD solutions. 
We have found a pair of real symmetry-broken ``underground'' solutions that have the correlation energy ranging from
$-265.333498\mbox{ eV}$ for $\beta=-5.0\mbox{ eV}$ to 
$-165.644985\mbox{ eV}$ for $\beta=-0.5\mbox{ eV}$. Moreover, for each $\beta$ there is a pair of
complex symmetric solutions (except for $\beta=-4.0\mbox{ eV}$, where there are
two pairs of such solutions)
with the real part of $E_{\rm corr}^{\rm CCD}$ lower than the ground-state CCD value.
For instance, for $\beta=-0.5\mbox{ eV}$ one has
$E_{\rm corr}^{\rm CCD}=-13.907202 \pm i 12.763757 \mbox{ eV}$, and for $\beta=-5.0\mbox{ eV}$
$E_{\rm corr}^{\rm CCD}=-3.892588 \pm i 110.045436 \mbox{ eV}$.
All the ``underground'' solutions do not resemble any of the FCI states;
such solutions have not been observed for H4 model.

Since the standard method of obtaining CCD solutions is via iterative
procedure~(\ref{eq:CCD3}), it may be useful to study the the performance of this procedure 
in attaining multiple solutions when starting from various CCD solutions obtained by the homotopy method.
Obviously, if the solutions were exact, 
the iterations would stop after the first iteration.
However, the small errors in the numerical values
would lead to a non-trivial iteration sequence.
The question is whether this sequence is convergent and, if yes, what is the converged result.
We tested all the real solutions and 
it is surprising that the series is either divergent or the result is identical to the
ground state. It means that {\em only\/} the ground state is stable
in the iterative process. It explains why the iterative process with the MP2 starting point
gives only the ground state solution.

A different behavior has been found when applying the direct inversion in the iterative subspace (DIIS) method~\cite{pulay:82,scuseria:86}.
It turned out that this method, which is useful for accelerating convergence, improves also the stability of the iterative process. 
Most of the states are stable in the DIIS iterative
process and only between 6--8 states out of the 22--32 real ones either diverge
or converge to a different state (not necessarily to the ground state). 
It is worth stressing that the ``underground'' solutions are iteratively unstable for all the cases tested.

\section{Concluding remarks}\label{sec:conc}

In this paper we have investigated the complete set of solutions of
the CCD equations corresponding to the PPP model of benzene. 
This is the largest system for which the complete set of solutions has been obtained. 
We have found that the number of solutions (for a broad range of the
resonance integral $\beta$) is surprisingly large (equal to 230),
approaching the limit given by the B\'{e}zout theorem (equal to 256).
One may wonder if some peculiar properties of the PPP model are
responsible for that proliferation of solutions. To this end, we
have checked the behavior of the PPP model of butadiene, which is
analogous to the H$4$ model studied previously by Jankowski and
Kowalski.~\cite{jankowski:99a,jankowski:99d} We have found, however,
that the CCD equations for these two models have roughly the same
number of solutions. We thus conclude that the large number of
solutions of the CCD equations corresponding to the PPP model of
benzene have to be related to some special properties of a cyclic
$\pi$-electron system. The number of solutions is expected to grow
exponentially for the larger cyclic polyenes (annulenes). The known
difficulties with solving the CC equations for these systems (see
Ref.~\onlinecite{podeszwa:02a} and references therein) are
undoubtedly a consequence of this multiple-solution problem. 
On the other hand, some non-standard solutions of the CCD equations for larger annulenes
may turn out to be similar to certain excited-state FCI solutions of these
systems.
Since attaining these solutions by means of the homotopy method seems
to be out of reach in the near future, we find it encouraging that the
DIIS method~\cite{pulay:82,scuseria:86} of carrying the
iterations in the CC method proved quite effective in assessing the
stability of several multiple solutions to the CCD equations for benzene.
This has inspired us to look for multiple solutions of the CCD
equations for the next member of the annulene family,
cyclodecapentaene (C$_{10}$H$_{10}$). Indeed, several new solutions
were found by combining a sort of random generation of the initial
$t_2$ amplitudes with the DIIS iterations. The results of these
investigations will soon be published.

\section*{Acknowledgements}
The work was supported in part by the Committee for Scientific Research (KBN) through Grant No. 7 T09A  019 20. 
The authors are grateful to Dr. Karol Kowalski for rending them access to his homotopy program and for valuable advice.


\end{document}